\documentclass[twocolumn,showpacs,amsmath,amssymb,prb]{revtex4}

\usepackage{graphicx}
\usepackage{dcolumn}
\usepackage{bm}

\begin{document}

\preprint{ver. 2.1}

\title{Magnetic phase diagram of antiferro-quadrupole ordering in HoB$_2$C$_2$}
\author{Tatsuya Yanagisawa}
\altaffiliation[Also at ]{Institute of Pure and Applied Physical Sciences, University of California at San Diego, La Jolla, California 92093, USA.}%
\author{Terutaka Goto, Yuichi Nemoto}
\affiliation{Graduate School of Science and Technology, Niigata University, Niigata 950-2181, Japan}
\author{Ryuta Watanuki}
\affiliation{Institute for Solid State Physics, University of Tokyo, Kashiwa, Chiba 277-8581, Japan.}
\author{Kazuya Suzuki}
\affiliation{Graduate School of Engineering, Yokohama National University, Yokohama 240-8501, Japan}
\author{Osamu Suzuki and Giyuu Kido}
\affiliation{National Institute for Materials Science, Sakura, Tsukuba 305-0003, Japan}
\date{\today}

\begin{abstract}
 The magnetic phase diagram for antiferro-quadrupole (AFQ) ordering in tetragonal HoB$_2$C$_2$
has been investigated by measurements of elastic constants $C_{11}$, $C_{44}$ and $C_{66}$ in fields along the
basal $x$-$y$ plane as well as the principal [001]-axis. The hybrid magnet (GAMA) in Tsukuba Magnetic
Laboratory was employed for high field measurements up to 30 T. The AFQ phase is no
longer observed above 26.3 T along the principal [001] axis in contrast to the relatively
small critical field of 3.9 T in fields applied along the basal [110] axis. The quadrupolar intersite
interaction of $O_{xy}$ and/or $O_2^2$ is consistent with the anisotropy in the magnetic phase
diagram of the AFQ phase in HoB$_2$C$_2$.
\end{abstract}

\maketitle
\begin{table*}[ht]
\caption{\label{tab:01}Classification and summary of the phase on HoB$_2$C$_2$. }
\begin{ruledtabular}
\begin{tabular}{lcr}
Phase&Properties&Reference\\
\hline
I & Para-magnetic, Para-quadrupole & \onlinecite{Onodera99,Yanagisawa03}\\
II & Antiferro-quadrupole ordering (with field induced magnetic moments)&(\onlinecite{Tanaka99,Yamauchi02,Tanaka04,Igarashi_Nagao03,
Lovesey01,Hirota00,Matsumura02_1})\footnotemark[1],\onlinecite{Matsumura02_2}, \onlinecite{Shimada01}\footnotemark[2]\\
III & Antiferro-quadrupole+Antiferro-magnetic ordering (coexistence)& \onlinecite{Ohoyama00,Matsumura02_2},\onlinecite{Zaharko04}\footnotemark[1],\onlinecite{Shimada01}\footnotemark[2]\\
III' & Antiferro-quadrupole+Antiferro-magnetic ordering (with magnetic domains)&\onlinecite{Zaharko04}\footnotemark[1]\\
III'' & Antiferro-quadrupole+Antiferro-magnetic ordering (with partly reoriented magnetic moments)&\onlinecite{Yamauchi99_1}\footnotemark[1],\onlinecite{Indoh04}\footnotemark[1]\\
IV & Incommensurate magnetic ordering (with diffuse neutron scattering)&\onlinecite{Ohoyama00,Tobo02}\\
\end{tabular}
\end{ruledtabular}
\footnotetext[1]{Results on DyB$_2$C$_2$.}
\footnotetext[2]{Results on Ho$_{1-x}$Dy$_x$B$_2$C$_2$.}
\end{table*}

\section{\label{sec:1}INTRODUCTION}
\begin{figure}
\includegraphics[width=0.60\linewidth]{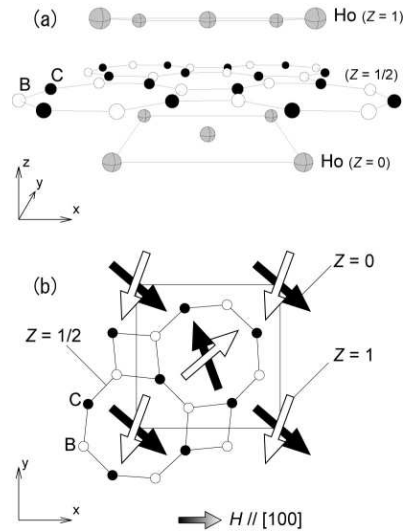}
\caption{\label{fig:00}
The crystal structure of HoB$_2$C$_2$. (b) The magnetic structure applying $H \parallel $[100] in phase III
of HoB$_2$C$_2$, according to Ohoyama {\it et al.}~\cite{Ohoyama00} and Zaharko {\it et al.}~\cite{Zaharko04}
}
\end{figure}

4$f$-electron systems, with orbital as well as spin degrees of freedom, in rare earth compounds
frequently show electric quadrupole ordering in addition to magnetic dipole ordering at low temperatures.
We noted that in HoB$_6$, with a $\Gamma_5$ triplet ground state, ferro-quadrupole ordering is
accompanied by a structural change from a cubic lattice to a trigonal one.\cite{Segawa92,Nakamura94,Goto00} CeB$_6$, with 
a $\Gamma_8$ quartet ground state, is well known as a typical example of antiferro-quadrupole
(AFQ) ordering in a cubic compound.\cite{Effantin85,Sakai97,Link98} The tetragonal lanthanide compounds DyB$_2$C$_2$,
HoB$_2$C$_2$ and TbB$_2$C$_2$ are also known to show AFQ ordering in competition with
antiferro-magnetic (AFM) ordering.\cite{Yamauchi99_1,Onodera99,Kaneko00}
These systems have the tetragonal LaB$_2$C$_2$-type structure~\cite{Yamauchi99_2,Duijn00,Ohoyama01} with space group P4/mbm
as shown in Fig.~\ref{fig:00}(a).
The specific heat and magnetic susceptibility measurements
 performed by Yamauchi et al. revealed that DyB$_2$C$_2$ exhibits quadrupole ordering of the
Dy$^{3+}$ $(J = 15/2)$ ions in phase II below $T_Q = 24.7$ K, which successively transits to an
AFM state in phase III at $T_N = 15.3$ K.\cite{Yamauchi99_1} Neutron diffraction measurements have shown the
characteristic AFM structure, with a slightly tilted angle, lying in the tetragonal basal plane.\cite{Duijn00,Yamauchi99_2}
The tilting of magnetic moments in the phase III are attributed to the competitive inter-site
interactions between the AFM and AFQ moments. Furthermore, field induced AFM moments
of the AFQ phase II in the basal plane were detected by neutron scattering under
magnetic fields applied along the [100]-axis.\cite{Yamauchi02}
Recently, Tanaka et al. performed detailed resonance x-ray scattering measurement at two resonant energies
of electric dipole (E1) transition and electric quadrupole (E2) transition at the Dy$^{3+}$ L$_{\rm III}$ absorption edge.
\cite{Tanaka99,Tanaka04} 
The (0 0 $l/2$) reflections at E2 channels of scatteting suggests the anisotropic charge distribution below $T_Q$ is consistent with the AFQ ordering in basal plane.  
It was also reported that an anisotropic charge distribution due to a small buckling lattice distortion of the B- and C- atoms contributes to the main peak
of the resonant x-ray spectra of E1 transition in DyB$_2$C$_2$.
\cite{Hirota00,Lovesey01,Matsumura02_1,Matsumura02_2,Igarashi_Nagao03}
These results suggest that the $O_{xy}$- and/or
$O_2^2$-type quadrupole ordering possessing a charge distribution in the basal plane is the order
parameter of phase II of DyB$_2$C$_2$.

\begin{figure}
\includegraphics[width=0.60\linewidth]{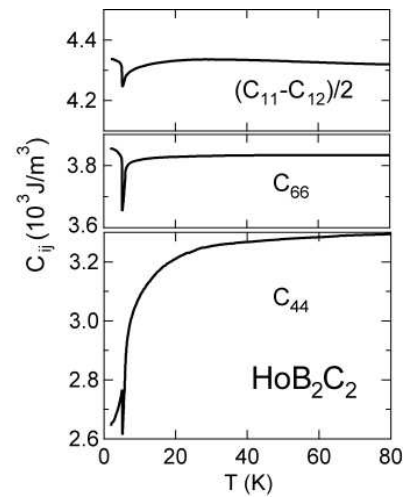}
\caption{\label{fig:01}Elastic constants of $C_{44}$, $C_{66}$ and $(C_{11}-C_{12})/2$ as a function of temperature below 80 K.}
\end{figure}

Onodera et al. reported that an isomorphous compound HoB$_2$C$_2$ shows an incommensurate
short range magnetic ordered phase IV at $T_{C1} = 5.9$ K and successively undergoes an AFM
ordering at $T_{C2}$ = 5.0 K in zero field.\cite{Onodera99}
The magnetic structure below $T_{C2}$ on HoB$_2$C$_2$ is essentially the same as that of
DyB$_2$C$_2$ (Fig.~\ref{fig:00}(b)).
The intermediate phase IV possesses a long periodic
magnetic structure characterized by a propagation vector of $k = (1+\delta , \delta , \delta ')$
with $\delta = 0.11$ and $\delta ' = 0.04$, along with broad diffuse magnetic scattering around $k = (100)$.
\cite{Ohoyama00,Tobo02}
The phase IV of HoB$_2$C$_2$ is contrast to the absence of that in DyB$_2$C$_2$.
It is noted that neutron scattering on TbB$_2$C$_2$ and ErB$_2$C$_2$ also show long periodic
magnetic ordering with nearly the same periodicity as HoB$_2$C$_2$.\cite{Kaneko02,Ohoyama02}
We show, in Table~\ref{tab:01}, a dyad of the phase classification and these properties on HoB$_2$C$_2$.
Here, phase I is para-magnetic (para-quadrupole) phase. The phase III' and III'' are sub-phase
of phase III, which will be discussed later. Some reference with superscript indicated
that the phase identifying by analogy from the results on DyB$_2$C$_2$ and Ho$_{1-x}$Dy$_x$B$_2$C$_2$ .
Elastic constants representing quadrupole susceptibility of the 4$f$-electron system is a useful
probe for examining a ground state with orbital degeneracy or pseudo-degeneracy in particular.\cite{Thalmeier_Luthi91}
We have performed ultrasonic measurements on HoB$_2$C$_2$ and in Fig.~\ref{fig:01} we present the measured
elastic constants for comprehension.\cite{Yanagisawa03} Considerable softening of 22 \% for $C_{44}$ below
100 K, 5.5 \% for $C_{66}$ below 50 K, and 2.4 \% for $(C_{11}-C_{12})/2$ below 30 K down to
$T_{C1} = 5.0$ K indicate a pseudo triplet ground state consisting of E-doublet and A (or B)
singlet in HoB$_2$C$_2$.\cite{Yanagisawa03} In phase IV, all transverse modes show enhanced softening
associated with considerable ultrasonic attenuation, where slow relaxation time of
$\tau \sim 7\times10^{-9}$ s was found.

The magnetic field-temperature ($H$-$T$) phase diagrams of DyB$_2$C$_2$ and HoB$_2$C$_2$ show
anisotropic behavior depending on the field directions.\cite{Onodera99,Indoh04} The order parameter of the AFQ
phase and its relation to the anisotropy of the $H$-$T$ phase diagram still remain to be solved.\cite{Yamauchi01}
The $H$-$T$ phase diagram of the AFQ phase II of tetragonal DyB$_2$C$_2$ and HoB$_2$C$_2$
compounds are often compared to that found in cubic CeB$_6$ \cite{Effantin85,Sakai97,Link98} and La-diluted systems
Ce$_x$La$_{1-x}$B$_6$.\cite{Hiroi97,Tayama97,Suzuki98} It has been reported that Ce$_x$La$_{1-x}$B$_6$ ($x = 0.75 \sim 0.60$)
exhibits an ordered phase IV being closely to the AFQ and AFM phases.
A huge softening in a transverse elastic constant $C_{44}$ and a trigonal lattice distortion
in the phase IV of Ce$_x$La$_{1-x}$B$_6$ indicates a spontaneous ferro-quadrupole moment.\cite{Akatsu02}
Kubo and Kuramoto have recently proposed a plausible model based on octupole ordering
to explain the trigonal distortion in phase IV.\cite{Kubo_Kuramoto03} The order parameter of phase IV in the
present compound HoB$_2$C$_2$ is not settled yet. The competition between AFQ and AFM
ordering in the tetragonal HoB$_2$C$_2$ system is an important issue in the present work.

We have performed ultrasonic measurements on single crystals HoB$_2$C$_2$ under
magnetic fields in order to investigate the anisotropic behavior of the AFQ phase II in the
$H$-$T$ phase diagram in fields along the three principal axes $H \parallel $ [110], $H \parallel $ [100] and $H \parallel $ [001].
In Sec. II the experimental procedure is described. The experimental results of the elastic constants and the magnetic phase
diagrams are presented in Sec. III. The concluding remarks are in Sec. IV.

\section{\label{sec:2}EXPERIMENTAL DETAILS}
HoB$_2$C$_2$ single crystals were grown with a tetra-arc furnace. Rectangular samples
with dimension of $3\times3\times2$ mm$^3$ and $2\times2\times3$ mm$^3$ were prepared
by a wire discharge cutter for the ultrasonic measurement.
The orientation of crystal with respect to applied magnetic field was settled with in the accuracy of 1 degree
by using x-ray Laue diffractions.
The LiNbO$_3$ transducers for the generation and detection of the sound waves with the frequencies
8.5 MHz and its overtone 31 MHz were bonded on opposite surfaces of the sample.
An ultrasonic apparatus based on the phase comparison method, detecting time-delay
for successive ultrasonic echo signals, was used to measure the sound-velocity $v$. For the estimation of
the elastic constant $C = \rho v^2$, we used the mass density $\rho  = 6.965$ g/cm$^3$ from
the lattice parameter $a = b = 0.534$ nm and $c = 0.352$ nm of HoB$_2$C$_2$. A $^3$He-evaporation
refrigerator was used for the low-temperature measurements down to 0.5 K.
Magnetic fields up to 12 T were applied by a superconducting magnet. Magnetic fields above 12 T
were generated by a hybrid magnet (GAMA) consisting of the superconducting magnet and
water-cooled resistive magnet in Tsukuba Magnet Laboratory (TML) of the National Institute
for Materials Science (NIMS).

\section{\label{sec:3}RESULTS AND DISCUSSION}
\subsection{\label{sec:3-A}Magnetic field dependence\\ for $H \parallel $ [100] and $H \parallel $ [110]}
\begin{figure}
\includegraphics[width=0.60\linewidth]{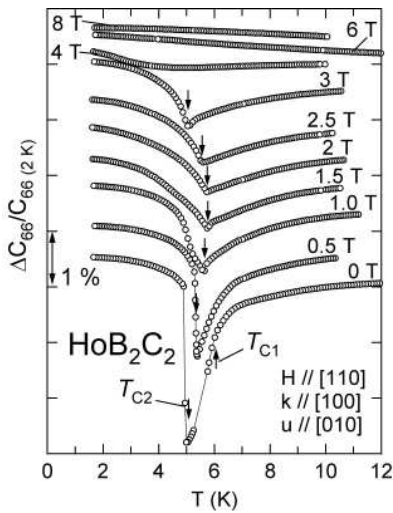}
\caption{\label{fig:02}Relative change of the elastic constant $\Delta C_{66}/C_{66}$ vs. temperature
under various magnetic fields applied along the [110]-axis of HoB$_2$C$_2$. Transverse modes at frequencies of 31 MHz
were used for the measurements.}
\end{figure}
Figure~\ref{fig:02} represents the relative change of elastic constant $C_{66}$ as a function of temperature
under various magnetic
fields for $H \parallel $ [110]. The $C_{66}$ was measured by the transverse ultrasonic mode
propagating along the [100]-axis with polarization along the [010]-axis, corresponding to a symmetry
strain $\varepsilon_{xy}$ with a B($\Gamma _2$) representation. 
The transverse $C_{66}$ exhibits a softening of 3 \% with decreasing temperature down to the
transition temperature $T_{C2} = 5.0$ K in zero field. The softening of $C_{66}$ in the phase IV
is suppressed in a magnetic field of 0.5 T, and $T_{C1}$ of the I-IV transition point shifts to lower
temperatures with increasing fields. The $C_{66}$ increases below $T_{C2}$ being the transition
point from the phase IV to the phase III. Then these I-IV and III-IV transition points cross
each other at a tetra critical point in $H$-$T$ phase diagram (See Fig.~\ref{fig:05} (c)). The minima of $C_{66}$
in fields above 1.5 T and up to 3 T indicates transition from paramagnetic phase I to the
AFQ phase II. Above 4 T, no indication of this phase transition was observed.
\begin{figure}
\includegraphics[width=0.60\linewidth]{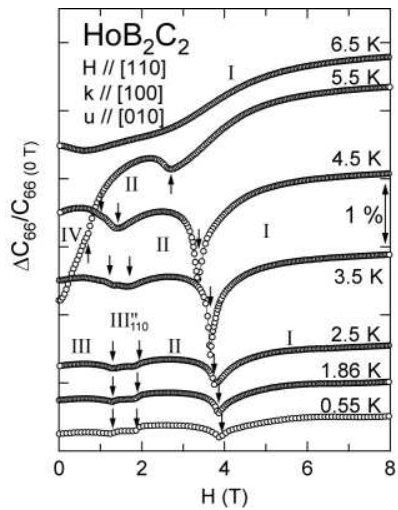}
\caption{\label{fig:03}Relative change of the elastic constant $\Delta C_{66}/C_{66}$ vs. magnetic field
at frequencies of 31 MHz under various temperatures in HoB$_2$C$_2$.
Magnetic fields up to 8 T were applied along the [110]-axis.}
\end{figure}
\begin{figure}
\includegraphics[width=0.60\linewidth]{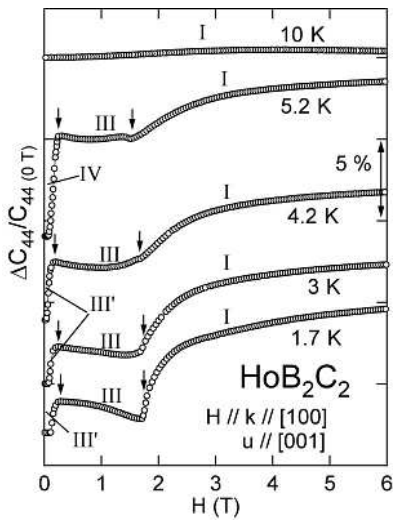}
\caption{\label{fig:04}Relative change of the elastic constant $\Delta C_{44}/C_{44}$ vs. magnetic field
at frequencies of 31 MHz under various temperatures in HoB$_2$C$_2$.
Magnetic fields up to 8 T were applied along the [100]-axis.}
\end{figure}

Figure~\ref{fig:03} represents the relative change of the elastic constant $C_{66}$ as a function of magnetic
field for $H \parallel $ [110] at several temperatures. In temperatures below 4.5 K and down to 0.55 K,
the sharp minima around 3.5 $\sim$ 4 T suggest the occurrence of an I-II phase transition, which is
expected in the ordered phase with symmetry breaking character. The intermediate region of
two transitions, from 2 T to 4 T in $H \parallel $ [110], indicates the AFQ phase II. The small
anomalies at fields lower than 1.9 T are an indication of the II-III phase transition. The III''$_{110}$-III
phase transitions has been also observed at 1.7 T. A broad minimum at 6.5 K indicates no sign
of the field induced phase transition. 

The relative change of the elastic constant $C_{44}$ as a function of magnetic field for $H \parallel $
[100] is shown in Fig.~\ref{fig:04}. The $C_{44}$ was measured by a transverse ultrasonic mode propagating
along the [100]-axis (or [010]-axis) with polarization along the [001]-axis. The $C_{44}$ mode
induces a symmetry strain $\varepsilon _{zx}$ (or $\varepsilon _{yz}$) corresponding to one
component in E($\Gamma _{34}$) doublet. The magnetic field dependence of $C_{44}$ in Fig.~\ref{fig:04}
shows two anomalies, a kink at 1.8 T corresponding to the I-III phase transition and a step anomaly
at 0.2 T. The later transition shows hysteretic behavior, which may be caused by a domain effect in
phase III. 
\begin{figure*}
\includegraphics[width=0.7\linewidth]{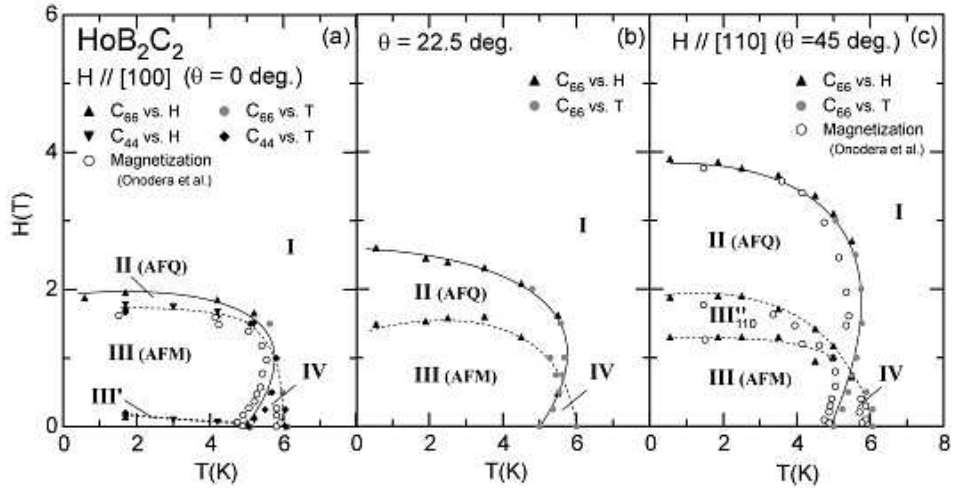}
\caption{\label{fig:05}$H$-$T$ phase diagrams of HoB$_2$C$_2$ with the fields applied along the (a) [100],
(b) $\theta  = 22.5$ deg. and (c) [110] directions. Data points were determined from the elastic anomalies in
$C_{66}$ vs. $H$ (solid triangles), $C_{66}$ vs. $T$ (grey circles), $C_{44}$ vs. $T$ (solid reverse triangles) and $C_{44}$ vs. $H$ (solid diamonds) as shown in Figs.~\ref{fig:02}-~\ref{fig:04}.
Solid and dotted lines are guide for eyes. Previous data of magnetization measured by Onodera et al. are also shown for comparison.}
\end{figure*}

In Fig.~\ref{fig:05}, we show the $H$-$T$ phase diagram of HoB$_2$C$_2$ which was obtained by
ultrasonic and magnetization measurements in fields applied parallel to the basal plane. Phase
boundaries were determined by $C_{66}$ vs. $H$ (solid triangles), $C_{66}$ vs. $T$ (grey circles), $C_{44}$
vs. $H$ (solid diamonds) and $C_{44}$ vs. $T$ (solid reverse triangles). 
Open symbols show the results of magnetization measurements by Onodera et al.\cite{Onodera99}
In the phase diagram of Fig.~\ref{fig:05}(a) we use the previous ultrasonic results from Ref. \onlinecite{Yanagisawa03}. In order to
verify the anisotropy of the AFQ phase II, the magnetic fields were applied along the intermediate
direction with an angle $\theta = 22.5$ deg. between the [100] and the [110]. Further details of
the ultrasonic results are skipped in the present paper for convenience.

As one can see in Fig.~\ref{fig:05}, the AFQ phase II shows anisotropic behavior as a function of the field
direction in the basal $x$-$y$ plane. The upper critical magnetic field $H_{C[110]} = 3.9$ T of the II-I
transition for $H \parallel $ [110] shrinks to $H_{C[100]} = 2.0$ T for $H \parallel $ [100],
while the AFM phase III and the phase IV behave mostly in an isotropic manner being independent
of the field direction in the basal plane. The order parameter of phase II has stability against field
$H \parallel $ [110] more than $H \parallel $ [100]. Actually, phase II is stable only in the vicinity
of the phase III-I boundary along $H \parallel $ [100] in Fig.~\ref{fig:05}(a).

\subsection{\label{sec:3-B}Magnetic field dependence for $H \parallel $ [001]}
\begin{figure}
\includegraphics[width=0.9\linewidth]{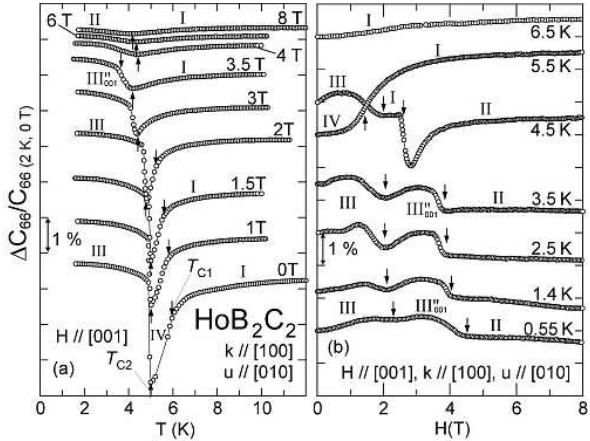}
\caption{\label{fig:06}Relative change of the elastic constant $\Delta C_{66}/C_{66}$
of the transverse modes at frequencies of 31 MHz under various fields and temperatures in HoB$_2$C$_2$.
Fields up to 8 T were applied along the [001]-axis. (a) shows temperature dependence, (b) shows field dependence.}
\end{figure}

\begin{figure}
\includegraphics[width=0.8\linewidth]{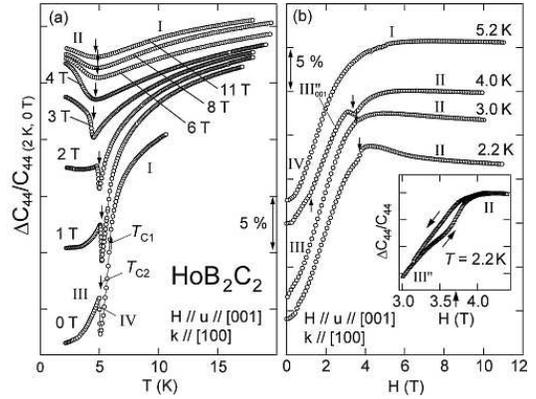}
\caption{\label{fig:07}Relative change of the elastic constant $\Delta C_{44}/C_{44}$ of the transverse modes at frequencies
of 31 MHz under various fields and temperatures in HoB$_2$C$_2$. Fields up to 11 T were applied along the [001]-axis.
(a) shows temperature dependence, (b) shows magnetic field dependence. Inset of (b) shows hysteretic behavior
around 3.5 K.}
\end{figure}

Figures~\ref{fig:06} (a) and (b) represent the temperature dependence of $\Delta C_{66}/C_{66}$
under various fields up to 8 T along the principal [001]-axis and field dependence at various
temperatures, respectively. In Fig.~\ref{fig:06} (a), the I-IV transitions at $T_{C1} = 5.9$ K, indicated by
down arrows, shift to lower temperatures with increasing field up to 3 T while the IV-III
transition $T_{C2}$, indicated by upward arrows, shift slightly to lower temperatures in field.
Eventually $T_{C1}$ and $T_{C2}$ cross each other at 3 T.  At 3.5 T, two anomalies indicating the
successive transitions I-II and II-III were observed. The I-II transition has been found from 4 T to 8 T
in Fig.~\ref{fig:06} (a). These transition points are shown in the phase diagram in Fig.~\ref{fig:10}. We successfully
observed the AFQ phase II in high magnetic field applied along $H \parallel $ [001] above 4 T.

Magnetic field dependence of $\Delta C_{66}/C_{66}$ at temperatures from 0.55 K to 6.5 K is shown in Fig.~\ref{fig:06}(b).
Two successive transitions of III- III''$_{001}$ and III''$_{001}$-II, indicated by arrows, are a common feature in the
measurements performed at 0.55, 1.4, 2.5, and 3.5 K. At 4.5K, a re-entrant process of III-I-II phase
transition was observed between 2 T to 2.7 T. At 5.5 K, the I-IV phase boundary shows a broad
plateau around 1.4 T. At 6.5 K, $C_{66}$ shows a monotonous increase in phase I. It should be noted
that the tetra-critical point exists at $T^* \sim 4.5$ K and in a field $H^* \sim 3$ T applied along the principal [001]
direction. This point is argued again in the magnetic phase diagram of Fig.~\ref{fig:10}.
\begin{figure}
\includegraphics[width=0.8\linewidth]{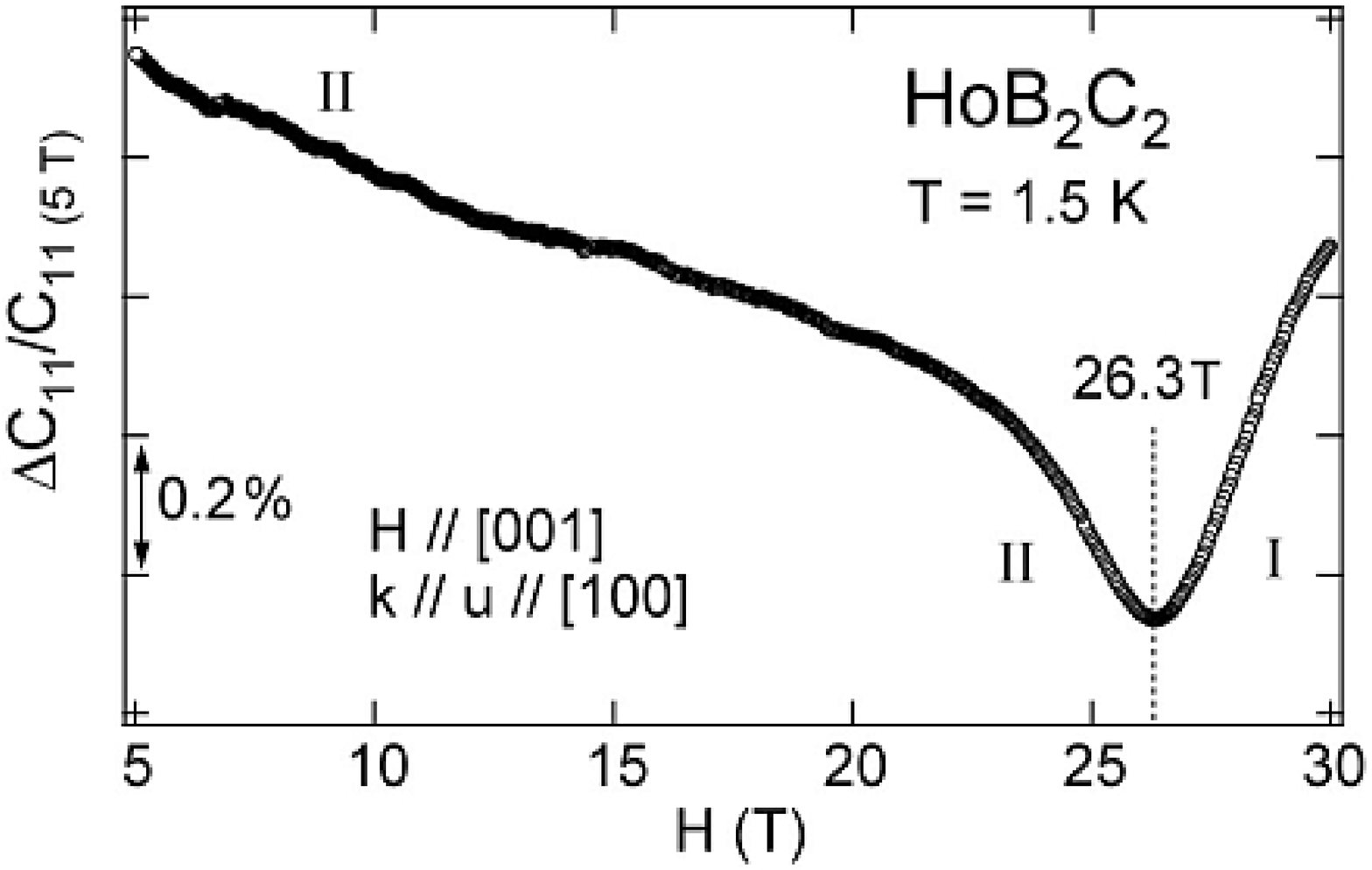}
\caption{\label{fig:08}Relative change of the elastic constant $\Delta C_{11}/C_{11}$
at a fixed temperature of 1.5 K displayed as a function of magnetic field along the [001]-axis up to 30 T. 
Measurement frequencies are 52 MHz.}
\end{figure}

As shown in Fig.~\ref{fig:01}, the transverse $C_{44}$ mode in zero field shows considerable softening on the
order of 20 \% with decreasing temperature. The softening of $C_{44}$ is very much reduced in applied
fields along the principal [001] direction, as shown in Fig.~\ref{fig:07}(a). The sharp minima of $C_{44}$ in fields below 3 T,
shown by downward arrows, indicate the IV-III transition points. The I-IV transition points,
which have clearly been observed in the results of $C_{66}$, were not identified in the results
of $C_{44}$ in Fig.~\ref{fig:07}(a). The shallow minima of $C_{44}$ above 4 T up to 11 T correspond to the
transition from the paramagnetic phase I to the AFQ phase II. 
In Fig.~\ref{fig:07}(b), we show magnetic field dependence of $\Delta C_{44}/C_{44}$ at various temperatures. The results
of 2.2, 3.0, and 4.0 K show the III-II transition around 4 T, indicated by arrows. As shown in the inset
of Fig.~\ref{fig:07}(b), a hysteresis effect has been observed which suggest the first order class of the III-II
transition. As will be shown in the phase diagram of Fig.~\ref{fig:10}, phase IV changes to the phase I with
increasing fields at 5.2 K. 

The magnetic field dependence of $\Delta C_{44}/C_{44}$ up to 11 T with $H \parallel $ [001] in Fig.~\ref{fig:07}(b) revealed
the field induced I-II transition. In order to examine the II-I phase transitions in higher fields with
$H \parallel $ [001], we have pursued ultrasonic measurements of $C_{11}$ using the hybrid magnet
(GAMA) up to 30 T in Tsukuba Magnet Laboratory. We chose the longitudinal $C_{11}$ mode because
of its definite ultrasonic echo signal as compared to the relatively faint echo signal in transverse ultrasonic
modes. Figure~\ref{fig:08} represents the magnetic field dependence of $\Delta C_{11}/C_{11}$ at 1.5 K for $H \parallel $ [001] from 5 T to 30 T.
A sharp minimum at 26.3 T has been found. The behaviors in AFQ phase transition of $C_{11}$
are very similar to the ones of $C_{66}$ for $H \parallel $ [100] or [110].
\begin{figure}
\includegraphics[width=0.60\linewidth]{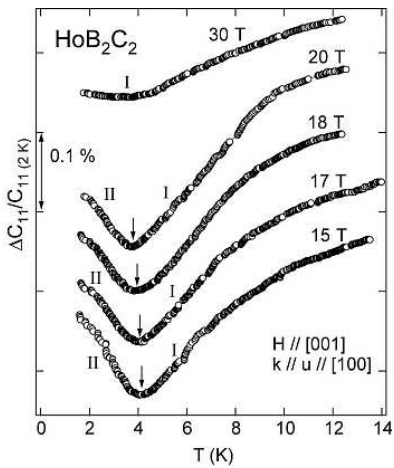}
\caption{\label{fig:09}Relative change of the elastic constant $\Delta C_{11}/C_{11}$ as a
function of temperature at frequencies of 52 MHz under various field applied along the [001]-axis.}
\end{figure}

Figure~\ref{fig:09} represents the temperature dependence of $\Delta C_{11}/C_{11}$ under fields of 15, 17, 18, 20 and 30 T.
The $C_{11}$ shows minima corresponding to the I-II phase transition, indicated by arrows,
around 4 K and below 20 T. The minima shift slightly to lower temperatures with increasing fields.
No anomaly in the result at 30 T indicates the absence of the I-II transition. The distinct difference
between the results of 20 T and 30 T suggests that phase II closes in the vicinity of critical field 
$H_{C[001]} = 26.3$ T.

Figure~\ref{fig:10} represents the magnetic phase diagram in field along the principal [001] -axis up to 30 T.
The gray circles, solid squares, solid triangles, solid reversed triangles and solid diamonds are the present results
of $C_{66}$ and $C_{44}$. The open triangles and open reversed triangles represent the transitions in
the present results of $C_{11}$ obtained by using the hybrid magnet. There are three ordered phases,
the AFQ phase II, AFM phase III, and phase IV in addition to the paramagnetic phase I.
In the phase III, sub-phase III''$_{001}$ is exist between 2 T and 4 T.
The magnetic neutron scattering in sub-phase III'' on HoB$_2$C$_2$ has not been reported yet.
On the analogy of a similar sub-phase III' on DyB$_2$C$_2$, the magnetic structure of sub-phase III''$_{110}$
for $H \parallel $[110] on HoB$_2$C$_2$ is expected the altered form of AFM structure in phase III,
which the magnetic moments are partly rotated to the advantageous direction for the external magnetic field.~\cite{Indoh04} 
On the other hand, the boundary of phase III''$_{001}$ for $H \parallel $[001] has not been
observed by magnetization measurement on HoB$_2$C$_2$.~\cite{Onodera99} Therefore, the origin of phase
III''$_{001}$ would be a different from phase III''$_{110}$.

As one can see in inset of Fig.~\ref{fig:10} , the four phases
meet each other at the tetra-critical point $T^* \sim 4.5$ K and $H^* \sim 3.0$ T.
It is notable that the I-II and IV-III phase boundaries approach the tetra-critical point tangential to each other, due to the interplay of several order parameters.
Similar multi-critical phenomena appear on the $H$-$T$ phase diagram of anisotropic
AFM systems.~\cite{Bruice75,Rohrer77}
The hysteresis effect across the II-III phase boundary ensures the first order transition, and the
discontinuity of the elastic constants at the IV-III transition points may also indicate the first order
transition.
\begin{figure}
\includegraphics[width=0.7\linewidth]{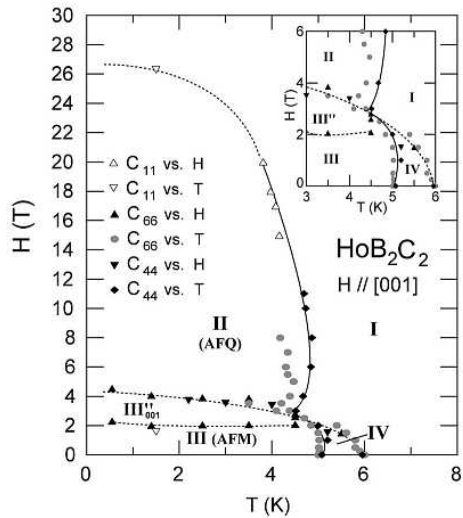}
\caption{\label{fig:10}Magnetic $H$-$T$ phase diagrams of HoB$_2$C$_2$ with fields applied along the [001]-axis.
Data points were determined from the elastic anomalies in
$C_{66}$ vs. $T$ (grey circles), $C_{66}$ vs. $H$ (solid triangles), $C_{44}$ vs. $H$ (solid reverse triangles),
$C_{44}$ vs. $T$ (solid diamonds), $C_{11}$ vs. $T$ (open reverse triangles) and $C_{11}$ vs. $H$ (open triangles)
as shown in Figs.~\ref{fig:06}-~\ref{fig:08}. Inset shows expanded view of tetra-critical point. Solid and dotted lines are guide for eyes.}
\end{figure}

The vertical I-II phase boundary from 4 T to 20 T is determined by the minima in temperature
dependence of $C_{44}$ and $C_{11}$. There is a difference between the data-plots $C_{66}$ and
$C_{44}$ that may be due to the mode difference or sample setting. We use the data of $C_{44}$
to obtain the I-II transition in high field. The point at 26.3 T obtained by $C_{11}$ in Fig.~\ref{fig:08} suggests an
upper phase boundary of the AFQ phase II in fields. The closed magnetic diagram of the AFQ phase in
the present HoB$_2$C$_2$ resembles the closed behavior in the AFQ phase II in Ce$_{0.5}$La$_{0.5}$B$_6$.\cite{Akatsu04}

\section{\label{sec:4}CONCLUDING REMARKS}
We have measured the elastic constants $C_{11}$, $(C_{11}-C_{12})/2$, $C_{44}$, and $C_{66}$ of HoB$_2$C$_2$.
The softening of these modes is due to the pseudo triplet ground state of the system.
The minima or kinks of $C_{11}$, $C_{44}$ and $C_{66}$ were useful in particular to
determine the transition points in fields. We have obtained the $H$-$T$ phase diagrams of HoB$_2$C$_2$
for the fields along the principal [001]axis and in the basal $x$-$y$ plane. In the $H$-$T$ phase diagrams,
there exists a tetra-critical point at $T^* \sim 5.5$ K, $H^* \sim 0.75$ T for the basal plane and
$T^* \sim 4.5$ K, $H^* \sim 3.0$ T for the principal [001]-axis, where two different interactions,
magnetic dipole and electric quadrupole, are competing with each other.

The AFQ phase II is stable even in fields of $H_{C[001]} = 26.3$ T along the [001] axis at absolute zero,
while the phase boundary shrinks to be $H_{C[110]} = 3.8$ T for fields along the [110]-axis and
$H_{C[100]} = 2.0$ T for [100]-axis. This anisotropy in the critical fields
$H_{C[001]} \gg  H_{C[110]} > H_{C[100]}$ in HoB$_2$C$_2$ is dominated by the anisotropy of
the quadrupolar RKKY-type quadrupole inter-site interaction mediated by the conduction electrons in the tetragonal
system. The de Haas-van Alphen measurements by Watanuki et al. revealed the main
Fermi-surface to have a columnar shape indicating two-dimensional character of the system.~\cite{Watanuki02}
The band calculation of LaB$_2$C$_2$ also shows the two-dimensional properties, reflecting the strong
hybridization of the $5d$ orbitals of La with the $2p$ orbitals in B-C sheets.~\cite{Harima00}
The anisotropic band structure will play a role in the anisotropic quadrupole inter-site interaction,
which brings about the anisotropic behavior of the AFQ phase of HoB$_2$C$_2$ and DyB$_2$C$_2$
under magnetic fields.

It is useful to demonstrate the symmetry properties of the quadrupole moments under the applied
fields along the high symmetry [100]-, [110]- and [001]-axis.
One may applied the symmetry argument of AFQ order parameters in cubic CeB$_6$, which was proposed by
Shiina et al.\cite{Shiina97,Shiina98,Sakai03}, to the present tetragonal HoB$_2$C$_2$ system. For a finite
magnetic field, the symmetry of the system is lowered to keep the field induced dipole moment, namely
angular momentum $J_x, J_y$ and $J_z$ to be invariant.
In the case of CeB$_6$, due to the $\Gamma _5$-type AFQ order parameter, linear combinations
of quadrupole moments result in cubic symmetry. In an applied magnetic field, the highest symmetry axis for these moments
are related to field directions as follows; $O_{xy}$ for $H \parallel $ [001], $O_{yz}+O_{zx}$ for $H \parallel $ [110]
and $O_{yz}+O_{zx}+O_{xy}$ for $H \parallel $ [111] in cubic symmetry O$_h$ .

In the case of tetragonal HoB$_2$C$_2$, when a field is applied along the [001]-axis,
the local symmetry of the rare earth ion of C$_{4h}$ is lowered to C$_4$ symmetry,
while fields applied [100]- or [110]-axes change C$_{4h}$ to C$_2$.
Since the [001]-axis of $\Gamma _2$-type quadrupole $O_{xy}$ and/or $O_2^2$
has a highest symmetry of three principal [100]-, [110]- and [001]-axis, this causes no modulation in
the $\Gamma _2$-type quadrupole when $H \parallel $ [001]. As a consequence, the $\Gamma _2$-type
AFQ order parameter $O_{xy}$ and/or $O_2^2$ would be stuck on the basal plane even in magnetic field
parallel to the basal plane due to the tetragonality of the system. 

\begin{acknowledgments}
The authors would like to thank J. Igarashi, O. Sakai, R. Shiina, and H. Onodera for fruitful discussions.
The present work was supported by a Grant-in-Aid for Scientific Research Priority Area "Skutterudite"
(No. 15072206) of the Ministry of Education, Culture, Sports, Science and Technology.
T. Yanagisawa was supported by Research Fellowships of the Japan Society for the Promotion of Science
for Young Scientists in present research.
\end{acknowledgments}

\end{document}